\documentclass[fleqn,usenatbib]{mnras}

\usepackage{newtxtext,newtxmath}

\usepackage[T1]{fontenc}
\usepackage{ae,aecompl}

\usepackage{graphicx}	
\usepackage{amsmath}	
\usepackage{amssymb}	

\title[Shock Breakouts from Tidal Disruption Events]{Shock Breakouts from Tidal Disruption Events}

\author[A. Yalinewich et al.]{
A. Yalinewich ,$^{1}$\thanks{E-mail: almog.yalin@gmail.com}
J. Guillochon$^{2}$
R. Sari,$^{3}$
and A. Loeb$^{2}$
\\
$^{1}$Canadian Institute for Theoretical Astrophysics, 60 St. George St., Toronto, ON M5S 3H8, Canada\\
$^{2}$Harvard-Smithsonian Center for Astrophysics, The Institute for Theory and Computation, 60 Garden Street, Cambridge, MA 02138, USA\\
$^{3}$Racah Institute of Physics, the Hebrew University, 91904, Jerusalem, Israel
}

\date{Accepted XXX. Received YYY; in original form ZZZ}

\pubyear{2015}

\begin{document}
\label{firstpage}
\pagerange{\pageref{firstpage}--\pageref{lastpage}}
\maketitle

\begin{abstract}
Tidal disruption events of stars by supermassive black holes have so far been discovered months to years after the fact. In this paper we explore the short, faint and hard burst of radiation is emitted at maximum compression, as a result of shock breakout. The detection of this burst can be used to capture tidal disruption events in real time. We verify that shock breakout from main sequence stars produces radiation in the X-ray range, but find that it is difficult to detect using all sky X-ray surveying telescopes. In the case of shock breakout from red giants, most of the radiation is emitted in the UV and visible range, which is significantly easier to detect. A similar burst of UV/optical radiation will also be emitted by stars puffed by tidal heating from a previous passage close to the central black hole. This radiation can be detected by surveys like ZTF and LSST. We calculate detection rates for different types of galactic nuclei. For the case of a very full loss cone we predict a detection rate of once per month with LSST, whereas for the case of a very empty loss cone we predict a rate of once per year with LSST. Evidence from a recent tidal disruption event, ASASSN-14li, seems to favour a very full loss cone, in which case LSST is expected to detect one such event every month.
\end{abstract}




\section{Introduction}

A star that passes too close to a black hole will be ripped apart by the tidal forces exerted by the black hole. Tidal disruption events occur in galactic nuclei \citep{rees_1988}. Recently, a tidal disruption event candidate has even been detected in a globular cluster, suggesting a possible interaction with an intermediate mass black hole \citep{lin_stradler_et_al_2018}. 

As will be explained in detail in the next section, tidal disruption begins when the distance between the black hole and the star drops below a critical distance called the tidal radius. The time the star spends inside the tidal sphere (i.e. a sphere with tidal radius around the black hole) is of the order of the dynamical time of the star (hours for main sequence stars). Yet, radiation is produced on much longer time scales, ranging from months to years depending on the mass of the black hole \citep{guillochon_ramirez_ruiz_2015}. This means that all tidal disruption events observed so far were captured a months to years after the fact.

In this paper we discuss an emission of radiation that occurs while the star is still inside the tidal sphere. This emission is due to the radiative shock breakout due to the compression of the star in a direction normal to the plane of motion. The formation of a shock around the equatorial plane and its motion toward the poles has been shown in hydrodynamic simulation simulations \citep{1983A&A...121...97C, guillochon_et_al_2009}. The actual shock breakout occurs on a very thin layer close to stellar surface, and involves complicated interaction between the gas and the radiation field, so it cannot be analysed using the same numerical simulation. In this work we apply the formalism for shock breakout, which has been studied extensively for supernova explosions \citep{nakar_sari_newtonian_breakout_2010,katz_et_al_2010}, to calculate the luminosity, duration and spectra from radiative shock breakouts from tidal disruption events.

The plan of the paper is as follows, in Section \ref{sec:analytic} we derive the formulae for the luminosity, duration and light - curves for radiative shock breakouts from tidal disruption events. In section \ref{sec:observability} we give predictions for the rate at which we expect these events to be detected using present day and next generation instruments. Finally, we conclude in Section \ref{sec:conclusions}.

\section{Analytic Predictions} \label{sec:analytic}

\subsection{The Plunge}

We consider a star of radius $R_{\rm s}$ and mass $M_{\rm s}$ approaching a black hole of mass $M_{\rm bh}$. Tidal disruption commences when the distance between the star and the black hole drops below the tidal radius (sometimes called Hill's radius)
\begin{equation}
R_{\rm T} \approx R_{\rm s} \left(\frac{M_{\rm bh}}{M_{\rm s}} \right)^{1/3} \, . \label{eq:tidal_radius}
\end{equation}
We assume that the centre of mass of the star is moving on a parabolic orbit with periapse distance $r_{\rm p}$. We define the ratio between the tidal radius and the periapse distance as the penetration parameter
\begin{equation}
\beta = \frac{R_{\rm T}}{r_{\rm p}} \, . \label{eq:penetration_parameter}
\end{equation}
After the star enters the tidal radius, tidal forces from the black hole begin to dominate over thermal pressure and self gravity in the star. For this reason, each of the fluid elements in the star evolves as though it is orbiting the black hole along a different Keplerian orbit. 

Let us now consider two points on the two poles of the star. These are points at maximum distance from the plane of motion of the star's centre of mass. The trajectory of these points can be obtained by rotating the trajectory of the star's centre of mass by an angle $R_{\rm s}/\sqrt{R_{\rm T} r_{\rm p}}$ along the line connecting the black hole to the periapse \citep{sari_stone_loeb_2013,coughlin_et_al_2015}. These three Keplerian orbits (two poles and the centre of mass) intersect at the periapse. The velocity difference between every two trajectories is of the order of
\begin{equation}
\Delta v_{\rm p} \approx \frac{R_{\rm s}}{\sqrt{R_{\rm T} r_{\rm p}}} v_{\rm p} \approx \beta \sqrt{\frac{G M_{\rm s}}{R_{\rm s}}}
\label{eq:hemisphere_velocity_difference}
\end{equation}
where the velocity of the centre of mass at periapse is
\begin{equation}
v_{\rm p} \approx \sqrt{\frac{G M_{\rm bh}}{r_{\rm p}}} \, .
\label{eq:periapse_velocity}
\end{equation}
We therefore expect that upon arrival to the vicinity of the periapse, the kinetic energy is
\begin{equation}
E_{\rm kp}  \approx M_{\rm s} \Delta v_{\rm p}^2 \approx \beta^2 \frac{G M_{\rm s}^2}{R_{\rm s}} \label{eq:kinetic_energy}
\end{equation}
In this paper we model a shock breakout from a TDE as though it were a supernova explosion with energy $E_{\rm kp}$.

As the star compresses, its thermal pressure increases, up to a point when the pressure can halt its collapse. At this point the approximation that particles move on Keplerian orbits no longer holds. In order to obtain the compression we assume that the initial speed of sound (prior to the star's entry into the tidal radius) is the same order of magnitude as the escape velocity, and that the compression is isentropic. The width of the star in a direction perpendicular to the plane of motion at peak compression is therefore \citep{brassart_luminet_2008}
\begin{equation}
z_{\rm p} \approx \beta^{-\frac{2}{\gamma-1}} R_{\rm s} \label{eq:peak_compression}
\end{equation}
where $\gamma$ is the adiabatic index. The projected area of the star on the plane of motion does not change considerably \citep{1983A&A...121...97C}, so most of the compression is perpendicular to the plane of motion \citep{sari_stone_loeb_2013}.

\subsection{Shock Ascent and Breakout}

As we discussed in the previous section, after the star enters the tidal radius it begins to collapse in the vertical direction. This collapse is halted by the increasing thermal pressure. This process causes a shock wave to emerge from the midplane of the star (the intersection of the star with the plane of motion) and move away from the plane of motion. When the shock reaches the stellar atmosphere, it encounters a steeply declining density. In this environment the velocity scales as a power law of the local density $v_{\rm s} \propto \rho_{\rm a}^{-\mu}$ where $\mu \approx 0.19$ \citep{sakurai_1960}. The density at the stellar atmosphere is assumed to vary as $\rho_{\rm a} \propto x^{\omega}$, where $x$ is the distance to the stellar edge and $\omega$ is a constant. In the case of an adiabatic atmosphere $\omega = \frac{1}{\gamma-1}$. The prefactor for the stellar density distribution can be obtained from the conservation of mass
\begin{equation}
\rho_{\rm a} \approx \frac{M}{R^2 z_{\rm p}} \left(x/z_{\rm p}\right)^{\omega} \, . \label{eq:stellar_atmosphere}
\end{equation}
At early times, the shock front is still close to the stellar core, where the density is constant, and hence it propagates as a Sedov Taylor Von-Neumann explosion \citep{sedov1993similarity,taylor1950formation,von1941point}. The velocity of the shock inside the star is given by
\begin{equation}
v_{\rm s} \approx \sqrt{\frac{E_{\rm kp}}{M_{\rm s}}} \left(\frac{x}{z_{\rm p}}\right)^{-\mu \omega} \, . \label{eq:shock_velocity}
\end{equation}
The shock is expected to be radiation dominated \citep{nakar_sari_newtonian_breakout_2010}, and so the acceleration ends when the the shock front is close enough to the stellar edge that photons can diffuse out. Assuming the dominant scattering process is Compton scattering, the depth of the breakout shell is 
\begin{equation}
\frac{c}{v_{\rm s}} = \frac{\rho_{\rm a}}{m_{\rm p}} r_{\rm e}^2 x_{\rm bo} \label{eq:breakout_shell_equation}
\end{equation}
where $c$ is the speed of light, $m_{\rm p}$ is the proton mass (we assume that the gas at the stellar edge is hydrogen), $r_{\rm e}$ is the classical electron radius (so $r_{\rm e}^2$ is the Thomson cross section) and $x_{\rm bo}$ is the depth of the shell from which breakout occurs. Shock acceleration stops when $x<x_{\rm bo}$. Solving for $x_{\rm bo}$ yields
\begin{equation}
\frac{x_{\rm bo}}{R_{\rm s}} \approx \beta^{\frac{- \gamma + 2 \mu \omega - 2 \omega - 1}{\left(\gamma - 1\right) \left(- \mu \omega + \omega + 1\right)}} \left(\frac{R_{\rm s}^2}{r_{\rm e}^2} \frac{m_{\rm p}}{M_{\rm s}} \sqrt{\frac{R_{\rm s} c^2}{G M_{\rm s}}} \right)^{\frac{1}{- \mu \omega + \omega + 1}} \, .
\end{equation}

\subsection{Prompt Emission}

The energy of the breakout shell is
\begin{equation}
\frac{E_{\rm bo}}{G M_{\rm s}^2/R} \approx \beta^{\frac{- \gamma \omega - \gamma + \omega + 1}{\gamma \mu \omega - \gamma \omega - \gamma - \mu \omega + \omega + 1}} \left(\frac{R_{\rm s}}{r_{\rm e}}\right)^{\frac{- 4 \mu \omega + 2 \omega + 2}{- \mu \omega + \omega + 1}} \times \label{eq:breakout_energy}
\end{equation}
\begin{equation*}
\times \left(\frac{M_{\rm s}}{m_{\rm p}} \sqrt{\frac{G M_{\rm s}}{R_{\rm s} c^2}} \right)^{\frac{2 \mu \omega - \omega - 1}{- \mu \omega + \omega + 1}} \, .
\end{equation*}
This energy is release over the periapse passage time \citep{guillochon_et_al_2009}. Along the trajectory the star is stretched by a factor of $\sqrt{\beta}$ \citep{sari_stone_loeb_2013}, so the periapse passage time is
\begin{equation}
t_{\rm p} \approx \frac{\sqrt{\beta} R_{\rm s}}{v_{\rm p}} \approx \sqrt{\frac{R_{\rm s}^3}{G M_{\rm s}}} \left(\frac{M_{\rm s}}{M_{\rm bh}} \right)^{1/3} \, . \label{eq:periapse_passage_time}
\end{equation}
The bolometric luminosity is
\begin{equation}
L_{\rm bo} \approx \frac{E_{\rm bo}}{t_{\rm p}} \approx \frac{G^{3/2} M_{\rm s}^{5/2}}{R_{\rm s}^{5/2}} \frac{\sqrt[3]{M_{\rm bh}}}{\sqrt[3]{M_{s}}} \times \label{eq:breakout_luminosity}
\end{equation}
\begin{equation*}
\times \left(\frac{R_{\rm s}}{r_{\rm e}}\right)^{\frac{- 4 \mu \omega + 2 \omega + 2}{- \mu \omega + \omega + 1}} \beta^{\frac{- \gamma \omega - \gamma + \omega + 1}{\gamma \mu \omega - \gamma \omega - \gamma - \mu \omega + \omega + 1}} \left(\frac{M_{\rm s}}{m_{\rm p}} \sqrt{\frac{G M_{\rm s}}{R_{\rm s} c^2}} \right)^{\frac{2 \mu \omega - \omega - 1}{- \mu \omega + \omega + 1}} \, .
\end{equation*}
Determining the temperature is less straightforward. Depending on the velocity and density, the shock can be in one of four different temperature regimes \citep{nakar_sari_newtonian_breakout_2010}. At low velocities the shock is matter dominated and the temperature is given by
\begin{equation}
k T_{\rm md} \approx m_{\rm p} v_{\rm s}^2 \label{eq:matter_dominated_temperature}
\end{equation}
where $k$ is the Boltzmann constant. At higher velocities the shock becomes radiation dominated. In this regime the temperature is
\begin{equation}
k T_{\rm rd} \approx \rho_{\rm a}^{1/4} v_{\rm s}^{1/2} c^{3/4} h^{3/4} \, \label{eq:radiation_dominated_temperature}
\end{equation}
where $h$ is Planck's constant. At even higher temperatures the shock enters the photon starved regime. In this regime the shocked fluid elements cannot produce enough photons to reach blackbody thermal equilibrium before they are swept downstream. The temperature in this regime is given by \citep{katz_et_al_2010}
\begin{equation}
k T_{\rm ps} \approx m_e c^2 \left(\frac{m_{\rm p}}{\alpha m_e}\right)^2 \left( \frac{v_{\rm s}}{c} \right)^8 \label{eq:photon_starved_temperature}
\end{equation}
where $\alpha$ is the fine - structure constant. Finally, above a certain velocity threshold the temperature saturates due to pair production. This happens at a temperature of around $T_{\rm ppo} \approx 200 \, \rm keV$, although photons only escape once the shells cools and pairs annihilate, which happens at $T_{\rm ppt} \approx 50 \, \rm keV$.

\begin{figure}
\includegraphics[width=0.9\columnwidth]{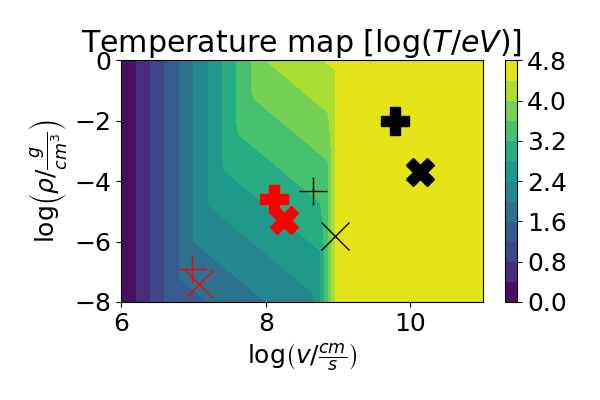}
\caption{
Temperature of the photons that diffuse out of the shock front as a function of velocity and density. The map also shows temperature, density and velocity for different scenarios. Red denotes red giants and black is main sequence. Convective envelopes by a plus sign, and radiative by an `x'. Markers with thin lines denote a shallow penetration $\beta=1$, while thick lines denote a deep penetration $\beta=10$.
}
\label{fig:temp_map}
\end{figure}

The different temperature regimes are summarised in figure \ref{fig:temp_map}. On the same figure we've also marked the corresponding effective temperatures for different kinds for breakout events. From this figure it is clear that only a red giant with a mild penetration factor can produce photons in the UV and visible range. Deep disruption of all kinds of stars and shallow disruption of main sequence stars produce most of the emission in the X ray range.

\section{Observability} \label{sec:observability}

\subsection{Main Sequence Star}
In this section we discuss the prospects of observing a shock breakout from a tidal disruption of a solar mass star from a $10^6 M_{\odot}$ black hole. We assume $\gamma = 5/3$, $\omega = 3/2$ and $\mu \approx 0.2$. 

Substituting these values into equation \ref{eq:breakout_luminosity} yields the luminosity of the breakout
\begin{equation}
L_{\rm bo} \approx 8.51 \cdot 10^{40} \tilde{M}_{\rm s}^{0.87} \tilde{M}_{\rm bh}^{0.33} \tilde{R}_{\rm s}^{-0.34} \beta^{1.14} \, \rm \frac{erg}{cm^3 s} \, .
\end{equation}
Where $\tilde{M}_{\rm s} = M_{\rm s}/M_{\odot}$, $\tilde{R}_{\rm s} = R_{\rm s}/R_{\odot}$ and $\tilde{M}_{\rm bh} \approx M_{\rm bh}/10^6 \, \rm M_{\odot}$.
The temperature of the photon is determined by pair production $T_{\rm ppt} \approx 50 \, \rm keV$. The duration of the burst is 
\begin{equation}
t_{\rm p} \approx 16 \tilde{R}_{\rm s}^{1.5} \tilde{M}_{\rm s}^{-0.17} \tilde{M}_{\rm bh}^{-0.33} \, \rm s \, .
\end{equation}

An X-ray telescope like NuStar \citep{ajello_et_al_2012}, with a sensitivity below $10^{-13} \, \rm erg/s/cm^2$, will be able to observe such events to distances of 90 Mpc. However, due to a small field of view and short duration of the event, the probability of a serendipitous discovery is negligible. A wide field sky survey in the x - ray like swift BAT has a much lower sensitivity - $10^{-8} \, \rm erg/s/cm^2$ \citep{barthelmy_2004}. For such a low sensitivity, a breakout from a main sequence star can only be detected from a distance of 200 kpc, i.e. from inside the milky way. We therefore conclude that such events cannot be observed with present day X-ray telescopes.

\subsection{Red Giants}

Part of the difficulties in observing main sequence stars are alleviated when considering the tidal disruption of red giants, due to their large radii. Assuming the radius of a red giant is around a hundred times larger than that of a main sequence star, then the time scales increase from a few seconds to a few hours, and the temperature of the photons drops from the x ray range to the UV / optical range. The downside is a reduction of the luminosity from $10^{41} \, \rm erg/s$ to $10^{40} \, \rm erg/s$. 

Compared to the rate of tidal disruption events of main sequence stars, the rate of tidal disruption of red giants decreases on the one hand because they are about ten times less common, but increases on the other hand because they have a radius that's about a hundred times larger \citep{macleod_et_al_2012}. The actual rate depends therefore on the type of loss cone. In the case of a full loss cone, the rate scales with the radius \citep{rees_1988}, so the overall rate of tidal disruption of red giants is about ten times larger than that of main sequence stars. In the case of an empty loss cone, the rate is logarithmic in the radius \citep{weissbein_sari_2017}, so overall the rate is about a factor of ten lower compared to main sequence stars.

In addition to ageing, tidal heating can also puff up stars. A star passing outside, but close to the tidal radius can be heated by the tidal force from the central black hole, such that the next time around its radius would be similar to that of a red giant \citep{li_loeb_2013}. In an empty loss cone the rate of disruption of tidally puffed stars is comparable to that of main sequence stars. In a full loss cone, the contribution of tidal puffing is negligible since a puffed star will most likely be scattered out of the loss cone before returning back to periapse. 

We consider the observation prospects with two optical surveying telescopes: ZTF \citep{bellm_kulkarni_2017} and LSST \citep{2017arXiv170801617R}. The limiting apparent magnitude of ZTF is about 20, with which events with luminosity $10^{40}$ erg/s can be detected up to a distance of about 17 Mpc. At this distance, the majority of galaxies, about 1000, lie in the Virgo cluster. LSST has a limiting magnitude of about 24, which allows it to detect the same events up to a distance of about 100 Mpc. At such a large distance we can use the average density of galaxies, roughly 0.01 per cubic Mpc, to calculate the total number of galaxies, about $5 \cdot 10^4$. The different estimates for the rate of tidal disruption events range between $10^{-4}$ and $10^{-5}$ per galaxy per year \citep{doi:10.1093/mnras/stv2281,van_velzen_2018}, and since they are closer to $10^{-4}$ per galaxy per year, we adopt it as our fiducial value. using these values we estimate in table \ref{tab:tde_rates} the expected detection rate for each scenario. The rates in the table also take into account the fact that both telescopes take between two to three nights to cover the whole sky, a period which is ten times larger than the duration of the event, which decreases the observed rate by a factor of ten. According to these rates, LSST is likely to detect at least one event within a ten year period. As for ZTF, in the most optimistic scenario it will detect a single event every decade. We note that, in theory, a dedicated survey with the same sensitivity as ZTF that only scanned the Virgo cluster each night might detect a single event every year, according to the most optimistic scenario.

Either a very full or a very empty loss cone (i.e. mean free time much smaller or much larger than orbital period) will give rise to the two most optimistic scenarios, for both of which we predict at least a single detection with LSST every year. There is, however, a possibility that a galactic centre will have a full loss cone for main sequence stars and an empty loss cone for red giants. This appears to be the situation for our galactic centre \citep{weissbein_sari_2017}. Moreover, in our galactic centre, red giants are under-represented, in comparison with galactic stellar population \citep{genzel_et_al_2010}, possibly due to red giant and stellar mass black hole collisions \citep{dale_davies_et_al_2009}. This does not necessarily affect the rate of tidal disruption events, as most originate from host galaxies different from the Milky Way \citep{french_arcavi_zabludoff_2016}. Data from one particular tidal disruption event, ASASSN-14li, suggest that the gas density around its galactic centre is about an order of magnitude larger than gas density around our galactic centre at similar distances \citep{krolic_et_al_2016,alexander_et_al_2016}. This might indicate that the stellar density is also larger by an order of magnitude, for which the loss cone might full for both main sequence and red giants.

One possible complication is extinction due to dust. In the Milky Way, the average galactic extinction due to dust is around 1.8 magnitude per kpc \citep{whittet2002dust}. Observations suggest that extinction can vary considerably between tidal disruptions \citep{auchettl_et_al_2017}, where some TDEs exhibit a negligible amount of extinction \citep{jiang_et_al_2016}, while other occur inside a layer of dust so thick it suppresses the visible and X-ray emission \citep{mattila_et_al_2018}. Since the estimates on the rates rely mostly on TDEs detected in the visible and X-ray range, in this study we neglect extinction.

\begin{table}
\caption{Expected TDE breakout detection rates, and the total expected number of detections with LSST.}
\label{tab:tde_rates}
\begin{tabular}{ p{1.8cm} | l l l }
Model                        & ZTF [yr$^{-1}$]       & LSST [yr$^{-1}$]  & LSST total    \\
\hline
red giant, full loss cone    & $10^{-1}$ & $10^{1}$ & $10^2$  \\
puffy stars, empty loss cone & $10^{-2}$ & $10^{0}$ & $10^1$      \\
red giant, empty loss cone   & $10^{-3}$ & $10^{-1}$ & $10^0$
\end{tabular}
\end{table}

\section{Conclusions} \label{sec:conclusions}

In this work we considered the emission from a radiative shock breakout close to maximum compression, in the course of a tidal disruption of a star by a black hole. We considered two types of stars: main sequence stars and puffed stars. The puffed stars include red giants and main sequence stars puffed by tidal heating from the black hole. For main sequence stars we expect a luminosity of about $10^{41}$ erg/s, a duration of about 20 seconds and an average photon energy of about 1 - 10 keV. In the case of puffed stars, the luminosity is about $10^{40}$ erg/s, a duration of a few hours and an average energy of a few dozen eV. Since the emission from main sequence stars is mostly in the X-ray range, and due to sensitivity of all sky X-ray surveys, we do not expect any such event will be discovered. On the other hand, the emission from puffed stars is emitted in the UV and visible range, where the instruments are more sensitive. We have considered several scenarios for the conditions at other galactic nuclei, and calculated the detection rates in each. In the most optimistic scenario we predict one discovery per century with ZTF. For LSST, our predictions range from once per year in the most optimistic case, to once per century in the pessimistic case. Circumstantial evidence from the tidal disruption event ASASSN-14li seems to favour the most optimistic scenario.

One channel that can boost the detection rates, but has not been considered in the previous sections is tidal disruption events from intermediate mass black holes in globular clusters \citep{2018arXiv180608385F}. It has been suggested that they occur at a comparable rate to tidal disruption events in galactic nuclei. Compared to a super - massive black hole, the radiative shock breakout from an intermediate mass black hole would produce a dimmer, but more prolonged emission. This trade - off is due to the fact that the total energy is independent of the mass of the black hole. The detection rate scales linearly with the duration, but super - linearly with the luminosity ($\propto L^{3/2}$ where $L$ is the luminosity), so even if tidal disruption occur at the same in globular clusters and galactic nuclei, it is the latter kind that would dominate the sample of observed shock breakouts.

Finally, we want to point out that very little is known about nuclei of other quiescent galaxies. Even a non detection of radiative shock breakouts within a certain period will tell us something about them. For example, a non detection within a decade with LSST will suggest that some process, like collisions with stellar mass black holes, destroys puffed stars. 

\section*{Acknowledgements}

AY would like to thank Chris Kochaneck, Wei Zhu and Amir Weissbein for the useful discussion. RS is supported by an ISF and an iCore grant. AL is supported in part by the Black Hole Initiative at Harvard University, which is funded by a JTF grant. In this work we made use of the NumPy \citep{oliphant2006guide}, SymPy \citep{10.7717/peerj-cs.103} and matplotlib \citep{hunter2007matplotlib} python packages.




\bibliographystyle{mnras}
\bibliography{almogyalinewich} 




\appendix

\section{Variable Glossary}

In the table below we summarise the meaning of each variable used

\begin{table} 
\centering
\caption{Variable glossary}
\label{tab:glossary}
\begin{tabular}{lllll}
 $R_{\rm s}$ & Star radius \\
 $M_{\rm s}$ & Star mass \\
 $M_{\rm bh}$ & Black hole mass \\
 $R_{\rm T}$ & Tidal radius (equation  \ref{eq:tidal_radius})\\
 $r_{\rm p}$ & Periapse distance\\
 $\beta$ & Dimensionless penetration parameter (equation \ref{eq:penetration_parameter})\\
 $\Delta v_{\rm p}$ & Velocity difference between stellar hemispheres (equation \ref{eq:hemisphere_velocity_difference}) \\
$v_{\rm p}$ & Centre of mass velocity at periapse (equation \ref{eq:periapse_velocity}) \\
$E_{\rm kp}$ & Kinetic energy of the star at periapse (equation \ref{eq:kinetic_energy}) \\
$z_{\rm p}$ & Width of the star at periapse (equation \ref{eq:peak_compression}) \\
$\gamma$ & Adiabatic index \\
$v_{\rm s}$ & Shock Velocity (equation \ref{eq:shock_velocity}) \\
$\rho_{\rm a}$ & Stellar atmosphere density (equation \ref{eq:stellar_atmosphere}) \\ 
$\mu$ & Sakurai parameter \citep{sakurai_1960} \\
$x$ & Distance to the stellar edge \\
$\omega$ & Stellar atmosphere density power law index \\
$c$ & Speed of light \\
$m_{\rm p}$ & Proton mass \\
$r_{\rm e}$ & Classical electron radius \\
$x_{\rm bo}$ & Depth of breakout shell (equation \ref{eq:breakout_shell_equation}) \\
$E_{\rm bo}$ & Energy of the breakout shell (equation \ref{eq:breakout_energy}) \\
$t_{\rm p}$ & Periapse passage time (equation \ref{eq:periapse_passage_time}) \\
$L_{\rm bo}$ & Breakout luminosity (equation \ref{eq:breakout_luminosity}) \\
$T_{\rm md}$ & Matter dominated temperature (equation \ref{eq:matter_dominated_temperature}) \\
$k$ & Boltzmann constant \\
$T_{\rm rd}$ & Radiation dominated temperature (equation \ref{eq:radiation_dominated_temperature}) \\
$h$ & Planck's constant \\
$T_{\rm ps}$ & Photon starved temperature (equation \ref{eq:photon_starved_temperature}) \\
$\alpha$ & Fine - structure constant \\
$T_{\rm ppo}$ & Saturation temperature due to pair production \\
$T_{\rm ppt}$ & Temperature at which pairs annihilate \\
$t$ & Time since breakout \\
$M_{\odot}$ & Solar mass \\
$R_{\odot}$ & Solar radius \\
$\tilde{M}_{\rm s}$ & $M_{\rm s}/M_{\odot}$ \\
$\tilde{M}_{\rm bh}$ & $M_{\rm bh}/10^6 M_{\odot}$ \\
$\tilde{R}_{\rm s}$ & $R_{\rm s}/R_{\odot}$
\end{tabular}
\end{table}

\bsp	
\label{lastpage}
\end{document}